# Au Nanocluster Growth on Graphene Supported on Ni(111)

Jory A. Yarmoff* and Christopher Salvo

*Department of Physics and Astronomy, University of California, Riverside, CA 92521*

**Abstract**

Low energy alkali ion scattering is used to investigate the deposition of Au onto a single layer of graphene grown onto Ni(111) by chemical vapor deposition. The yield of 3.0 keV $Na^+$ singly scattered from Au as a function of coverage indicates that it grows in a Volmer-Weber mode forming nanoclusters that increase in size with the amount of deposition. The neutralization probability of the scattered $Na^+$ is high for the smallest clusters and decreases as they increase in size. This is presumably caused by the cluster edge atoms being positively charged combined with the fact that the ratio of edge to center atoms decreases with size, which is similar to the behavior of Au nanoclusters on oxide substrates. In addition, oxygen is intercalated under the graphene film to decouple it from the substrate, but no changes in the growth mode or neutralization probability are observed.

*Corresponding author, E-mail: yarmoff@ucr.edu

# I. Introduction

Supported metal nanoclusters that consist of a few to thousands of atoms have many potential applications, particularly as catalysts [1-3]. Such nanoclusters can be grown by physical vapor deposition (PVD) from an atomic beam onto a surface with a higher free energy than the bulk metal so that the atoms form islands and do not wet the surface, which is the Volmer-Weber growth mode [4-6]. Metals atoms directly deposited onto oxide substrates typically grow fairly flat nanostructures that are only a few atomic layers thick [7,8], in contrast to gas phase clusters which are more three-dimensional [9]. The average size of the PVD-grown clusters increases with the amount of Au that is deposited.

Gold (Au) nanoclusters supported on oxide surfaces are particularly strong catalysts, and their activity is size dependent [2,10-12]. The adsorption of precursors and the subsequent catalytic reactions are generally considered to occur at the edges of the clusters [13-17]. In addition, density functional theory (DFT) has shown that the periphery atoms of Au clusters on oxide supports are positively charged while the center atoms are nearly neutral [13,14,18]. It has been proposed that oxygen in the substrate forms strong chemical bonds to the clusters that leads to charge transfer away from the edge atoms, and that this bonding is what produces the catalytically active sites on the nanoclusters [1,19]. Note that PtMo and PtRuMo clusters supported on highly-oriented pyrolytic graphite (HOPG) are also catalytically active, however, even in the absence of oxygen in the substrate [20].

There has also been much recent interest in novel two-dimensional 2D materials, such as graphene [21], single layer transition metal dichalcogenides [22], and topological insulators [23,24]. Graphene (Gr) is a particularly important material because of its unusually high electron mobility [25], tensile strength [26], thermal conductivity [27], and surface area to mass ratio [28].

The band structure of Gr contains a signature Dirac cone that gives rise to Dirac electrons making it an attractive material for device applications [21]. Films of Gr can be grown via chemical vapor deposition (CVD) onto various metal surfaces, including Ni(111) [29].

The PVD growth of metals onto Gr surfaces has been studied extensively [6,30]. Au and other noble metals readily diffuse across Gr leading to the formation of nanoclusters via a similar Volmer-Weber process as for metals deposited on oxides such as $TiO_2$ and $SiO_2$ [7,31]. For example, Au PVD forms clusters on Gr supported on Ir(111) [32] as well as on hydrogenated graphene, either free standing or supported on Cu [33], although it does not form nanoclusters on Gr films on Ru(0001) [30].

Nanoclusters supported on a Gr substrate are being explored for use as nanocatalysts [34]. As examples, Ru nanoclusters supported on Gr are highly active in the hydrolysis of ammonia borane [35] and Ni nanoparticles on Gr activate aldehyde hydrosilylation [36]. It has also been predicted that Pd nanoclusters on Gr can dissociate $H_2$ and store large amounts of hydrogen [37]. In addition, nanoparticles deposited onto modified Gr that is either oxidized, contains surface vacancy defects, is doped with N, or supports large molecules along with the nanoclusters are active nanocatalysts [38-43].

A novel form of alkali low energy ion scattering (LEIS) [44,45] is used here to probe Au nanoclusters deposited by PVD onto Gr films on Ni(111). LEIS is extremely surface sensitive [46] and able to monitor the growth mode and average charge state of the Au atoms in the nanoclusters [18]. LEIS employing 3.0 keV $Na^+$ projectiles shows that the deposited Au atoms do form nanoclusters on Gr/Ni(111). The neutralization probability as a function of the amount of deposited Au decreases with coverage, which is an indication that the atoms at the periphery of the clusters are positively charged [18]. In addition, oxygen is intercalated to increase the spacing between the Gr and Ni and decouple the film from the substrate [47], but no changes in the growth mode or the Na$^+$ neutralization probability are observed.

## II. Experimental Procedure

The experiments are performed in an ultra-high vacuum (UHV) chamber that has a base pressure of $1.2 \times 10^{-10}$ Torr. The chamber contains an xyz rotary manipulator and the instrumentation for low energy electron diffraction (LEED), x-ray photoelectron spectroscopy (XPS), and LEIS.

A commercially purchased Ni(111) 5x5x0.5 mm$^3$ polished single crystal (MTI Corp) is mounted to the foot of the manipulator and Ar$^+$ ion sputtering and sample heating are used for *in situ* sample preparation. The procedure involves cleaning the Ni(111) sample with cycles of 1.0 keV Ar$^+$ sputtering with a total beam current on the order of 1 µA for 1 hour and annealing at 700°C for 45 minutes. The cycles are repeated until a sharp 1x1 hexagonal pattern is observed by LEED. The formation of a pristine Ni(111) surface is further confirmed by both XPS and LEIS (data not shown).

Graphene is grown by CVD onto the clean Ni(111) surface by exposing it to 4500 L (1 L = $10^{-6}$ Torr sec) of ethylene while the sample is held at 600°C [47]. This is a self-limiting process and 4500 L is a sufficient exposure to ensure that a complete layer of graphene is produced. As a confirmation, XPS spectra were collected after 600, 2400, 4200, and 7800 L of ethylene exposure. The ratio of the C 1s to the Ni 2p core-level peaks increases until 4200 L at which point it saturates, thus confirming the formation of a complete Gr layer. The LEED pattern is unchanged by Gr growth, which is consistent with the presence of a single layer because the lattice mismatch of Gr with Ni(111) is only 1.2% [47,48]. In addition, LEIS spectra collected after the 4500 L ethylene exposure show no scattering from Ni indicating that the substrate is completely covered.

Further measurements are made by intercalating $O_2$ between the Gr and Ni(111) substrate prior to Au deposition [47]. The Gr/Ni(111) sample is exposed to $1.4 \times 10^5$ L of $O_2$ at 120°C. This creates a LEED pattern with additional spots (not shown) that are indicative of domains of Gr rotated with respect to each other, consistent with previous observations following oxygen intercalation that denote a decoupling of the Gr layer from the Ni substrate.

Au is deposited by PVD onto the Gr/Ni(111) and Gr/O/Ni(111) surfaces, as well as on bare Ni(111). The evaporator is composed of high purity Au wire (99.99%) wrapped around a tungsten filament (Mathis) that is heated to produce a thermal atomic beam of Au. This same procedure has been successfully used to grow Au nanoclusters on oxide substrates [7,49]. The Au evaporation rate is calibrated by a crystal quartz microbalance and reported in units of monolayers (ML), where 1 ML is defined as the number of Au atoms in single atomic layer of Au(111), which has a thickness of 2.6 Å [50].

Time-of-flight (TOF) is used to collect neutral and charged particle LEIS spectra [45]. 3.0 keV $Na^+$ ions are generated by a thermionic emission gun (Kimball Physics) and pulsed at 40 kHz with a pulse width of roughly 80 ns. The gun is mounted at a 30° angle from the TOF detector, leading to a 150° scattering angle. The detector is a triple microchannel plate (MCP) array located at the end of a 0.46 m long flight leg. A 1 cm aperture is located in front of the MCP, which defines an acceptance angle of 1.2°. The TOF leg contains a set of deflection plates used to separate ions from neutrals. When both plates are grounded, all of the scattered particles are detected, but when there is 300 V potential placed across the plates, only the scattered neutral particles reach the detector. Total yield and neutrals spectra are collected simultaneously by turning the voltage on and off every 60 seconds during the approximately 20 minutes it takes to collect the spectra, which eliminates any artifacts that could be caused by long-term drifts in the ion beam current. Ion scattering is an inherently destructive process, as each incoming ion acts to remove material from the surface. To minimize beam damage, the sample is re-prepared before the surface is impacted by a fluence that is equivalent to ~1% of a ML, which insures that the ions always probe pristine material.

### III. Results

The kinematics of the interaction of low energy ions with solids can be treated classically as a series of elastic binary collisions between the projectile and individual target atoms located at lattice sites since the differential scattering cross sections are smaller than the interatomic spacings [46]. Because the atomic bonding energies in solids are negligible compared to the projectile kinetic energy, the target atoms can be considered to be unbound so that the energy lost by the projectile following a collision is taken up by the recoiling target atom. Thus, each scattering event reduces the projectile kinetic energy by an amount that depends on the target atom to projectile mass ratio and the scattering angle. Note that there are additional small energy losses due to inelastic excitation processes [51], but these can generally be ignored in experiments such as those performed here.

The flight times of the scattered projectiles are converted to kinetic energy using the known projectile mass and flight tube length to generate a LEIS energy spectrum. Projectiles that experience one collision from a single surface atom and exit the sample directly into the detector produce single scattering peaks (SSPs) whose kinetic energy provides the elemental identification of the target atoms. The area of an SSP is proportional to the number of those surface atoms that are directly visible to both the ion beam and detector. Furthermore, the scattered projectiles are differentiated with respect to their charge state by collecting separate spectra for the "total" (ions and neutrals) and "neutral" yields.

Figure 1 shows typical TOF total and neutral yield LEIS energy spectra for 3.0 keV $Na^+$ scattered from 0.50 ML of Au deposited on Gr/Ni(111). The figure shows two similarly shaped spectra, the larger one being the total yield and the smaller one the neutral yield. The primary feature in each spectrum is the Au SSP, as 3.0 keV $Na^+$ elastically scattered at 150° from an isolated Au atom retains a kinetic energy of 1.9 keV. The spectra have the same basic shape following all Au coverages except that the size of the SSPs relative to the background changes. There are no other SSPs in the spectra despite the presence of C and Ni because 1) Na is more massive than C so that it cannot singly scatter at this large angle, and 2) the graphene film shadows the Ni substrate from the incident ion beam which precludes single scattering from Ni. Some of the emitted projectiles have also impacted multiple target atoms, which generates a multiple scattering background that spans a range of energies.

TOF-LEIS with alkali projectiles is surface sensitive largely because of shadowing and blocking. Shadow cones are regions behind a surface atom from which the ion beam is excluded due to scattering from that atom, which prevents single scattering from deeper lying atoms [52,53]. Blocking cones are formed when projectiles scatter from an atom located below the first atomic layer but cannot reach the detector due to another atom being positioned between the first target atom and the TOF leg. The surface sensitivity enabled by shadowing and blocking depends on the crystal structure and the ion incidence and exit directions. In previous work, a detailed analysis was used to show that low energy $Na^+$ singly scattered from PVD-grown Au nanoclusters probes only the outermost Au atoms in the clusters [18].

The number of Au atoms at the surface of the clusters is thus directly proportional to the area of the SSP. The normalized area of the Au SSP is presented as a function of the amount of Au deposited on Gr/Ni(111) in Fig. 2. The error bars are smaller than the size of the markers. To produce this figure, the Au SSP in each total yield spectrum is integrated after subtracting the multiple scattering background. Typical backgrounds are indicated by the dashed lines in Fig. 1. To ensure consistency across spectra, the integrated values are normalized by the amount of time over which the spectrum was collected and the average ion beam current. The error bars are calculated prior to normalization by taking the square root of all the counts under the SSP, including the background, which assumes that the error is purely statistical.

As Au is initially deposited, the total yield SSP signal increases linearly and the solid line in that portion of Fig. 2 is a fit of the data to a straight line. At the point marked by the vertical dashed line just below 1 ML in Fig. 2, the slope begins to curve downward. The solid line above this coverage is a fit of the data to a natural logarithm function. The change from a linear slope to a curved asymptotic lineshape indicates that the Au initially deposits in a single layer but when the slope changes the coverage is sufficient so that the Au begins to form multilayer clusters. The slope changes because some of the newly deposited atoms adsorb atop an existing Au atom and the spectra are only sensitive to the number of outermost atoms in the clusters so that the Au SSP area does not increase in direct proportion to the total Au coverage. The Au SSP area is not saturated even after deposition of 12 ML, indicating that a complete Au has not yet been formed. The behavior of the data in Fig. 2 directly shows that the Au grows in a Volmer-Webber mode, as often occurs in the PVD of metals on oxide supports [7,31].

The probability for neutralization of the Na$^+$ ions singly scattered from Au is measured and reported as the neutral fraction (NF) in Fig. 3. To determine the NF experimentally, the SSP of Na$^+$ scattered from Au in the neutral and total yield spectra are individually integrated after background subtraction, as described above, and the neutral SSP area is then divided by that of the total yield. The corresponding errors from each SSP area are propagated through the calculation to determine the error limits of the measured NFs.

The neutralization process for alkali projectiles scattered from metal surfaces is explained in detail elsewhere [54-56], and specifically for scattering from Au nanoclusters on TiO$_2$(110) in Ref. [18]. In brief, as an ion approaches a surface, the ionization level sees its image charge and shifts up in energy, while at the same time the level broadens by hybridizing with states in the solid [54,55]. The shifting and broadening are such that when the projectile is close to the surface, the broadened ionization level overlaps the Fermi energy of the solid and is resonant with the surface bands. When the projectile is close to the surface, electrons quantum mechanically tunnel between the ionization level and the solid leading to some amount of neutralization. This resonant charge transfer (RCT) process is slower than the scattering time, however, so that the neutralization occurs nonadiabatically. This means that the charge distribution of the scattered projectiles is frozen in along the exit trajectory when the projectile is at a distance from the surface at which electron tunneling becomes unlikely [57]. The measured NF thus depends on the position of the Fermi energy associated with the local electrostatic potential (LEP) directly above the target atom with respect to the shifted and broadened ionization level at this freezing point. The LEP in this context is sometimes referred to as the local work function. This process leads to an increase in the NF when the local work function decreases, and vice versa.

When scattering from a clean and flat metal surface, the NF is dependent upon the perpendicular component of the exit velocity, which is a function of the exit angle. When the exit trajectory is close to the surface plane, the NF will be higher than when it is normal to the surface because the perpendicular component of the exit velocity is reduced which decreases the non-adiabaticy of the neutralization process and thus increases the effective freezing distance [55,58]. This is not the case for nanoclusters supported on an oxide substrate, however, for which the NF remains constant as a function of emission angle due to the fact that the clusters are small and are thus more round than they are flat on the atomic scale [44]. A measurement of the NF as a function of exit angle for 0.20 and 0.30 ML of Au on Gr/Ni(111) is plotted in the inset of Fig. 3. The data show that the NF remains relatively constant as a function of emission angle, which further supports the conclusion that Au atoms deposited on Gr/Ni(111) form clusters similar to those on an oxide substrate.

Figure 3 shows the NF of the Au SSP for scattered 3.0 keV Na$^+$ as a function of the amount of Au deposited on three different samples. The samples are bare Ni(111) (triangles), Gr/Ni(111) (circles), and Gr/O/Ni(111) (squares) for which the oxygen is intercalated [47]. The NF for Na$^+$ scattered from Au decreases with coverage for all three samples. It is about 50% for the smallest depositions on Gr/Ni(111) and Gr/O/Ni(111) and reduces to about 3% after deposition of about 3 ML when the cluster sizes are large enough to have the same LEP as a complete Au film. The NF starts at about 20% for deposition on bare Ni and also reduces to about 3% after a 2 ML deposition. Note that a NF of about 50% is similar to that of Na$^+$ scattered from the smallest Au nanoclusters on oxide substrates, while the value of 3% is consistent with measurements from bulk Au [18,44].

Au deposited on clean Ni(111) is expected to deposit as a film and not form clusters as on an oxide or graphene. A confirmation of the absence of nanoclusters for Au/Ni(111) is that the Au SSP NF's are much smaller than those measured from Au nanoclusters. The change in the Au SSP NF with Au coverage from about 20% to 3% can be attributed solely to the differences in work function between Ni(111) and polycrystalline Au surfaces, which are 5.24 [59] and 5.40 eV [60], respectively. Although these values are fairly close to each other, such a work function increase in going from a Ni to a Au surface is sufficient to explain the accompanying decrease in NF with Au coverage.

The large initial NF values for small Au nanoclusters on Gr and the decrease with size are another confirmation that nanoclusters are formed. It is suggested in a recent paper that the decrease of the NF with Au coverage in scattering from nanoclusters on oxides is caused by the combination of having positively charged edge atoms and neutral center atoms in each cluster [18], as was predicted from DFT [13,14]. Ions that scatter from the edge atoms will have a higher NF than those that scatter from the center atoms because the positively charged periphery atoms form upward pointing dipoles that decrease the LEP above them. Since the experimentally measured NF is an average over all of the Au scattering sites, it is dependent on the ratio of the number of edge to center atoms. As the clusters grow in size, this ratio decreases so that the NF also decreases. Reference [18] provides a parametric model using this concept that closely matches experimental neutralization data for $Na^+$ and $Li^+$ scattering from Au clusters on $TiO_2(110)$, confirming this explanation for the neutralization of alkali ions scattered from nanoclusters. Thus, the decrease of NF with Au deposition on Gr is an indication that the cluster edge atoms in this system are also positively charged.

Gr strongly bonds to Ni(111) because of the close lattice match, and is thus void of a Dirac cone in its band structure [21]. When oxygen is intercalated between the graphene and Ni, two changes occur that could affect the supported nanoclusters. First, intercalated oxygen decouples the Gr from the Ni substrate which causes a change to the LEED pattern [47] and restores the Dirac cone associated with freestanding Gr [21]. Second, it physically increases the spacing between the Gr and the Ni substrate from 0.21 to 0.26 nm [47,61,62]. The intercalation of $O_2$ prior to Au deposition does not, however, alter the NF in Na$^+$ scattering, as shown by the data in Fig. 3. The independence of the NF on oxygen intercalation suggests that the coupling of Gr to the Ni and the interlayer spacing do not affect the growth of Au nanoclusters on Gr nor the charge associated with the edge atoms in the deposited clusters.

IV. Discussion

The data presented here show a reduction in the neutral fraction with cluster size, which is identical to the behavior of Au clusters on TiO$_2$ [44,63,64], SiO$_2$ [49,65], Al$_2$O$_3$ [66] and HOPG [66,67]. This behavior is consistent with the model that ascribes the reduction in NF with Au cluster size to the edge atoms being positively charged while the center atoms are neutral [18].

The assertion that the clusters contain positively charged atoms is supported by reports in the literature that examine core level shifts in high-resolution synchrotron x-ray photoelectron spectroscopy (SXPS) for small clusters on oxides [68,69] and graphite [70]. The SXPS spectra show both a shifting up in binding energy and a broadening of the metal core level for the smaller clusters. The shifting up was accredited to positive charge being present in the cluster and the broadening to final-state effects, but this broadening may also be, at least partially, a consequence of the inhomogeneity of the charge across each nanocluster.

The positively charged edge atoms in Au clusters on TiO$_2$(110), as initially reported by DFT calculations, suggest an interaction between the atoms at the periphery of each cluster and the bonding substrate O atoms [13,14]. This is not surprising since O is more electronegative than Au and would thus pull negative charge from the bonding Au atoms. C has about the same electronegativity as Au, however, so it may not be expected that the bonding Au atoms would be positive. The change of the LEIS NFs with cluster size suggest, however, that the edge atoms are indeed positively charged when they are resident on Gr. Therefore, an electronegative substrate atom is not the only way to create positively charged periphery atoms in a Au nanocluster.

In the present case, it is likely that the positively charged Au edge atoms are a result of their bonding to the vacant $\pi$ bonds of the Gr layer, which protrude upward from the surface [21]. Bonding to such $\pi$ orbitals could remove charge from the edge atoms in a similar manner as for Au bonding to a substrate O atom. The NF for low energy Li$^+$ scattered from Au nanoclusters supported on HOPG show a similar decrease of the NF as a function of Au coverage [66,67], which suggests that the bonding of the edge atoms to the substrate is via $\pi$ orbitals for both single layer Gr and bulk graphite surfaces.

It should be noted that PtMo and PtRuMo clusters deposited on HOPG are catalytically active [20]. Thus, an oxide substrate is not necessary to produce nanocatalysts. Although there has been no direct connection made between the positive charge and the catalytic activity, they are both associated with the undercoordinated edge atoms [16] and it has been proposed that this undercoordination is the reason for the nanocluster's catalytic ability [1]. Since nanoclusters on oxides, HOPG and Gr all have positively charged edge atoms, it can be inferred that that this charge plays a role in the catalytic activity whether the charge results from bonding to electronegative O atoms or to $\pi$ orbitals.

V. **Conclusions**

LEIS employing Na$^+$ ions is used to investigate the deposition of Au on a Gr film on a Ni(111) substrate. The yield of singly scattered projectiles as a function of the amount of Au deposited indicates that nanoclusters are formed. The neutralization of scattered Na$^+$ ions from the Au nanoclusters decreases with cluster size, which indicates that the edge atoms are positively charged presumably through bonding to surface π orbitals. No changes in the growth mode or charge state are observed when oxygen intercalation is used to decouple the Gr layer from the substrate showing that the properties of the nanoclusters depend on the interaction of Au atoms with Gr and are not influenced by the underlying substrate. The results of this work show that Au clusters form on Gr much in the same way as they do on oxide and graphite substrates, both in terms of their growth mode and their charge distribution. Since the positively charged periphery atoms are known to be the active sites on oxide substrates, they are presumably also the active sites for metal nanocatalysts supported on other materials.

## VI. Acknowledgements

The authors would like to thank Vivek Aji for useful discussions. This material is based upon work supported by the National Science Foundation under CHE - 1611563.

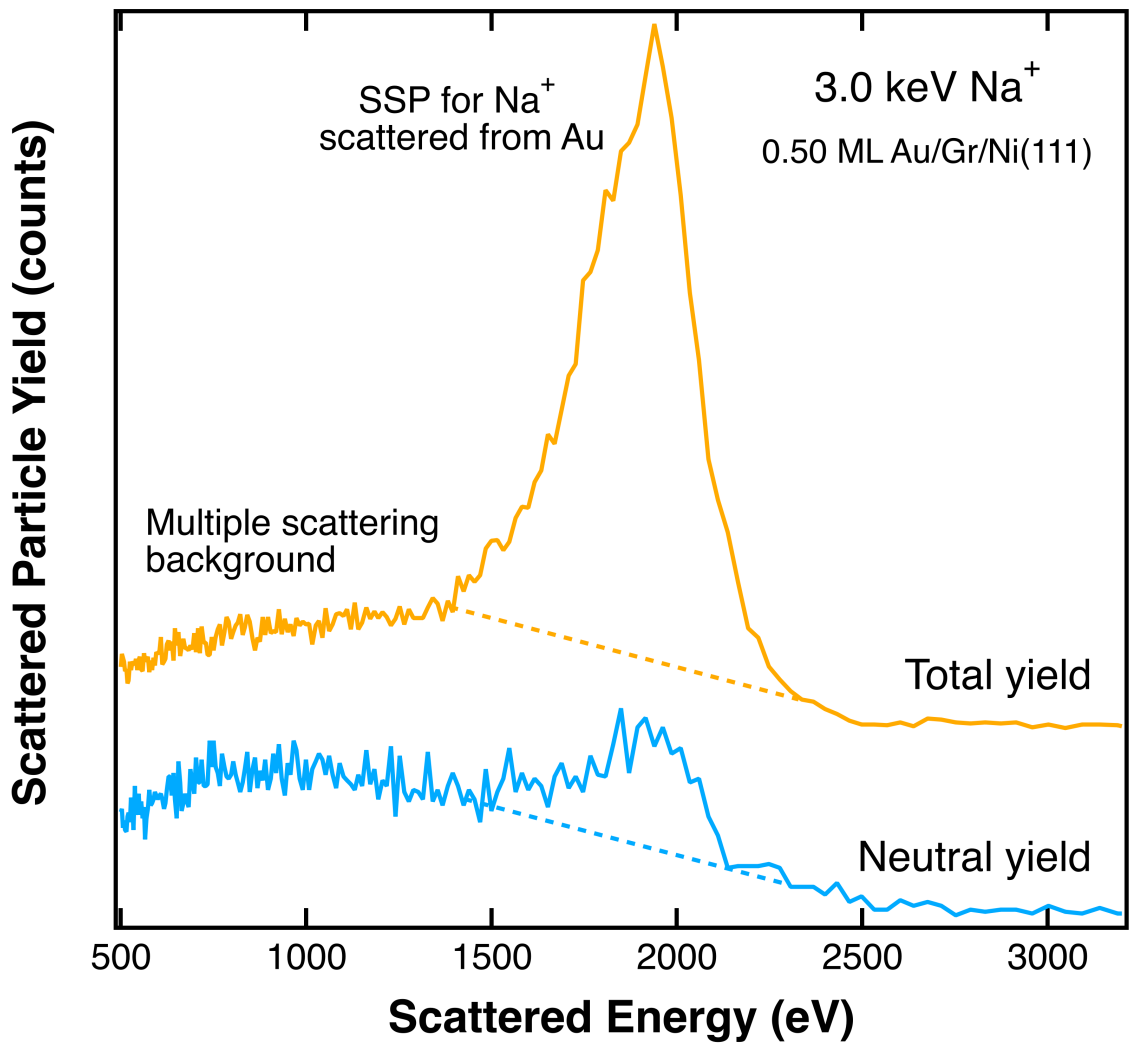

**Figure 1.** Typical TOF LEIS spectra of 3.0 keV Na$^+$ scattered at 150° from 0.50 ML Au deposited on a graphene film grown on Ni(111). The upper spectrum is the total yield and the lower spectrum is the neutral yield. The backgrounds are shown by dashed lines.

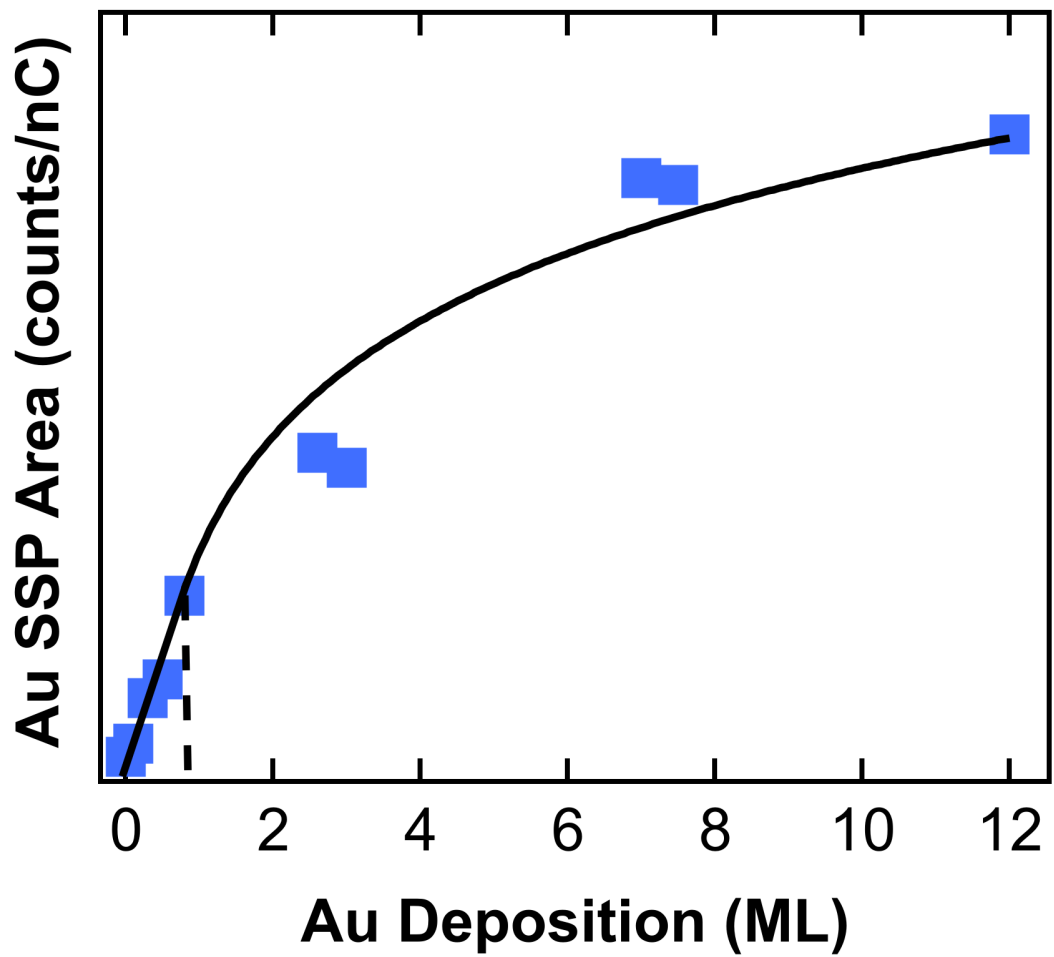

**Figure 2.** Total yield of 3.0 keV Na$^+$ scattered from Au as a function of Au deposition on Gr/Ni(111). The solid curve is a linear fit to the data up to the vertical dashed line and a logarithmic fit above it.

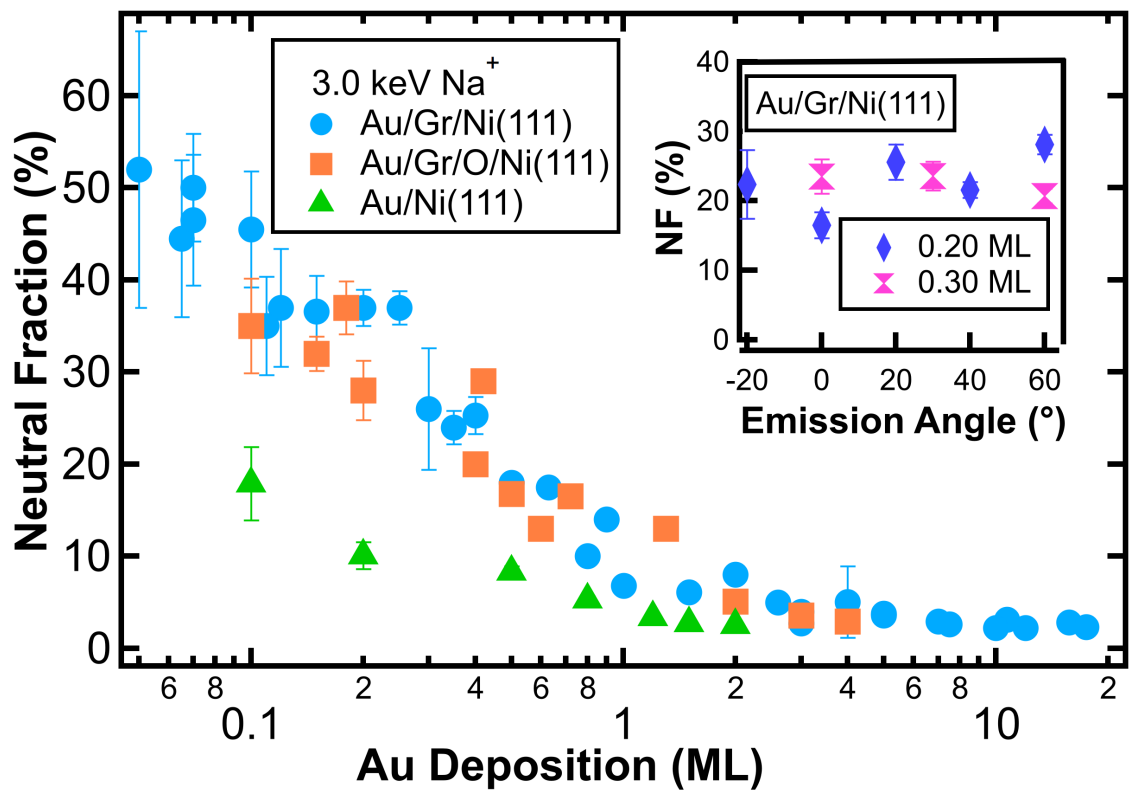

**Figure 3.** The Au SSP NF for 3.0 keV Na$^+$ scattered from Au as a function of Au deposition for 3 different samples: Au on graphene on Ni(111) (circles), Au on graphene on Ni(111) with intercalated oxygen (squares), and Au on Ni(111) (triangles). Inset: The NF of 3.0 keV Na$^+$ scattered from 0.20 (diamond) and 0.30 ML (double triangles) of Au on Gr on Ni(111) as a function of emission angle.